\documentstyle[11pt] {article}

\title{Energy Density of a Dissipative Polarizable Solid
 by a Lagrangean Formalism}
\author{R. Englman$^{a,b}$ and  A. Yahalom$^b$ \\
$^a$ Department of Physics and Applied Mathematics,\\
Soreq NRC,Yavne 81800,Israel\\
$^b$ College of Judea and Samaria, Ariel 44284, Israel\\
e-mail: englman@vms.huji.ac.il; asya@ycariel.yosh.ac.il;}

\begin{document}
\maketitle

\newcommand{\beq} {\begin{equation}}
\newcommand{\enq} {\end{equation}}
\newcommand{\ber} {\begin {eqnarray}}
\newcommand{\enr} {\end {eqnarray}}
\newcommand{\eq} {equation}
\newcommand{\eqs} {equations }
\newcommand{\mn}  {{\mu \nu}}
\newcommand{\sn}  {{\sigma \nu}}
\newcommand{\rhm}  {{\rho \mu}}
\newcommand {\SE} {Schr\"{o}dinger equation}
\newcommand{\sr}  {{\sigma \rho}}
\newcommand{\bh}  {{\bar h}}
\newcommand {\er}[1] {equation (\ref{#1}) }
\newcommand{\gb} {{\bf \gamma}}
\newcommand{\gcrb}  {{\bf \gamma^+}}
\newcommand{\gd} {{\dot \gamma}}
\newcommand{\gcr} {{\gamma^+}}
\newcommand{\gcrd} {{ \dot \gamma^+}}
\newcommand{\ro} {{ \gamma  \gamma^+}}

\begin {abstract}
A Lagrangean for the dynamics of an electromagnetic field in
a dispersive and dissipative material is constructed (adapting some ideas
 by Bekenstein and Hannay)
and an expression for the energy density that is positive
 is obtained from it. The expression contains extra (sink) degrees of freedom
 that represent dissipating modes. In simplified cases the sink modes
 can be eliminated to yield an energy density expression in terms of the
 electromagnetic fields, the polarization and the magnetization only, but which
contains parameters associated with the sink modes. The method
  of adding extra modes can be used to set up a Lagrangean formalism for
  dissipative systems in general, such that will reinstate time-translation
 invariance and will yield a unique energy density.

\end {abstract}
\section {\it PACS:} 03.50.De; 71.36.+c
\section {\it Keywords:} Electromagnetic energy; Lagrangean formalism; Variational Techniques;
 Dissipation; Polarizable solid
\section {Background}
The problematic nature of the electromagnetic field energy in a dissipative material
is apparent already at a graduate-teaching level: a frequently used
textbook notes that in a dispersive medium the energy density
lacks a thermodynamical interpretation \cite {LandauLP}. (A
dispersive material is necessarily dissipative, since by the
Kramers-Kronig relations its constitutive constants, that are by
definition frequency ($\omega$) dependent, must have an imaginary part, which
represents absorption or energy loss.) The expression for the
energy density that was derived in \cite {LandauLP}, was valid
only for fields that are nearly mono-chromatic. For other basic
treatments we quote textbooks as \cite {Stratton}-\cite
{Jackson}, and note that the last reference labels the expression
obtained under the above conditions as an {\it
effective} energy density.

 We write the relations between the electric field and the displacement
using the frequency  dependent permittivity $\epsilon (\omega)$
 and between the magnetic field and the induction using the permeability
 $\mu (\omega)$, as
\beq
\vec D (\vec x,\omega) =  \epsilon (\omega) \vec E (\vec x,\omega), \qquad
\vec B (\vec x,\omega) =  \mu (\omega) \vec H (\vec x,\omega).
\label{epsmu}
\enq
Then according to \cite{Jackson} we  obtain an effective energy density:
\ber
u_{eff} &=&
Re \left[ \frac{d (\omega \epsilon)}{d\omega}(\omega_0)\right]
\langle \vec E (\vec x,t) \cdot \vec E (\vec x,t)\rangle
\nonumber \\
&+& Re \left[ \frac{d (\omega \mu)}{d\omega}(\omega_0)\right]
\langle \vec H (\vec x,t) \cdot \vec H (\vec x,t)\rangle
\label{ujackson}
\enr
The brackets $\langle\rangle$ designate an averaging over a period of
$\frac{2 \pi}{\omega_0}$, where $\omega_0$ is the carrier frequency.

Alternative expressions that were subsequently proposed were
controversial. One derivation, that required a significant departure from
standard electromagnetic theory \cite {Neufeld66}, postulated the
independence of the expression of particular material constants. This was
strongly criticized in \cite {Loudon}, where the energy-density formula
contained explicitly  parameters that were present in the equation of
motion for the field. Another derivation, in terms of constitutive constants
\cite {AskneL}, was found to lead to energy densities that are negative
for a medium with a narrow resonance \cite{Ziolkowski}.

The quest for a non-controversial energy density can be regarded from a
more fundamental angle. This quantity is expected to be part of a
conservation equation, which (by a well-known theorem due to Noether) is tied
to the time invariance of the Lagrangean. As soon as the invariance is
lost (which is the case for dissipative systems), the quantity to be
conserved is undefined. Thus the need to find an appropriate energy density
is expected to arise in a wider context, too. (In a dramatic account, in which
the proposer of a time-varying light-velocity describes his Iliad to get his idea
 published, one also finds the problem of the proper Lagrangean
 formulation to play a key role \cite {Magueijo}.)

The issue of the electromagnetic-field energy density in a
dispersive and dissipative medium has recently resurfaced in the
contexts of the subluminality of light-propagation and of left
handed materials (\cite{GlasgowWP},\cite{Ruppin}). The former work
utilizes some analytical properties of the constitutive relations,
while the departure point in the latter are the equations of
motions for the macroscopic polarizabilities. As an application,
the energy density for a left handed medium has been calculated in
\cite {Ruppin}. The expressions in the two papers differ.

 Our approach is based on the recognition that in standard field theories
(classical or quantal) the energy momentum tensor satisfies a
conservation equation having  the form of a "div"-equation involving all
degrees of freedom. Thus, it seems, that one should be able to write out
an energy-density in an unambiguous fashion starting with a Lagrangean.
(The idea of a Lagrangean
 formulation was raised earlier \cite{Neufeld69}, but only in a
 programmatic manner and by making an assumption that the detailed
 treatment worked out in this paper does not justify.)

 As is well known,
the  energy density $T^0_0$ is  a component of the
energy momentum tensor $T^k_j$, which is uniquely derivable from the
Lagrangean density $ {\cal L}(u_k,u_{k,j})$, this being a function of the
field variables $u_k$ and of their  derivatives $ u_{k,j}$. (The indexes take the value
 $0$ for the time component and the numbers $(1,2,3)$ for the remaining,
 space components. Summation is implied for repeated
  indexes and a symbol after the comma represents a derivative. The dot will
 also be used for a time derivative.) The formal definition is given by
\beq
T^k_j = \frac{\partial {\cal L}}{\partial u_{i,k}}u_{i,j}-
  {\cal L}\delta ^k_j
\label {Tkj}
\enq

Two circumstances appear to prevent one from deriving the energy-density
from the conservation equation in a dissipative material. First, a Lagrangean has not been formulated for the
equations of motion (but this is done below). Secondly, the energy
sinks involved in the dissipative mechanisms have not been given a
dynamic representation, but only a phenomenological one  (see \er{epsmu}),
namely, through the
appearances of complex permittivity and permeability in the constitutive
relations or, equivalently, through the presence of time-reversal
 non-invariant terms in the equations of motion.

 The first issue was recently resolved in a short note \cite{Hannay}, where it
 was shown how to formulate the Lagrangean (actually, the Hamiltonian)
 for a dissipative case. The second point was (indirectly) addressed in
 \cite {Bekenstein2002}, following a previous publication in \cite{Bekenstein1982}.
  These works formulated the dynamics of a (conjectural) time dependent fine-structure
  constant by including in the Universe an additional degree of freedom.
  We follow up both these approaches, with changes required by the
  different context.

\section {A Lagrangean for static polarization and  magnetization}

Before approaching the general problem we will first present a Lagrangean density
for the electromagnetic field in a material which has a static polarization and  magnetization.
The following expression connects the electric displacement field $\vec D$ with the electric field $\vec E$
and polarization $\vec P$:
\beq
\vec D = \epsilon \vec E + \vec P
\label{DE}
\enq
($\epsilon$ is not necessarily the vacuum permittivity $\epsilon_0$). Similarly the magnetic
field $\vec H$ is connected with the magnetic induction field  $\vec B$ and magnetization $\vec M$ by:
\beq
\vec H = \frac{1}{\mu} \vec B - \vec M
\label{HB}
\enq
(again $\mu$ is not necessarily the vacuum inverse permeability $\mu_0$).
The above fields satisfy both the homogeneous Maxwell's equations:
\beq
curl \vec E + {\dot{\vec B}}=0, \qquad ~div\vec B=0
\label{Maxhom}
\enq
and the inhomogeneous equations:
\beq
curl \vec H - \dot{ \vec D}=\vec J, \qquad ~div \vec D= \rho
\label{Maxinhom}
\enq
In the following we assume that both free charges $\rho$ and currents $\vec J$ are zero.
The above equations can {\it not} be obtained from a Lagrangean density expressed
in terms of those fields.  However, this situation can be amended by representing
the fields in terms of vector $\vec A$ and scalar $\Phi$ potentials, as follows:
\beq
\vec E = -\vec \nabla \Phi - \dot{ \vec A}, \qquad \vec B = curl \vec A
\label{pot}
\enq
Using these definitions we see that the homogenous \eqs (\ref{Maxhom})
 are satisfied automatically. The inhomogeneous \eqs (\ref{Maxinhom}) can be obtained
 from the functional derivative of the Lagrangean:
\ber
{\cal L} &=& {\cal L}_{EB} + {\cal L}_{PE} + {\cal L}_{MH}
\\
{\cal L}_{EB} &=& \frac{1}{2}[\epsilon \vec E^2 - \frac{\vec B^2}{\mu}]
\\
{\cal L}_{PE} &=&  \vec P \cdot \vec E
\\
{\cal L}_{MH} &=&  \mu \vec M \cdot \vec H
\label{staticlag}
\enr
in which the reader should think of all field quantities as given in terms
of the potentials of \er{pot}. Using \er{Tkj} one can obtain the energy density:
\beq
 T^0_0 = \frac{\partial {\cal L}}{\partial \dot{\vec A}}\cdot \dot{\vec A}-
  {\cal L}.
  \label{edk1}
  \enq
which yields for the Lagrangean of \er{staticlag} after some manipulations the
result:
\beq
 T^0_0 = \frac{1}{2}[\epsilon \vec E^2 + \mu \vec H^2 + \mu \vec M^2]
  \label{edstatic}
 \enq
The last term $\mu \vec M^2$ can be usually discarded for static magnetization
since it represents nothing but a time-independent constant. However, we include it for
future sections in which the polarization and magnetization will become dynamic
 degrees of freedom.

  \section {Equations of motion for polarization and magnetization}

In cases that the polarization ${\vec P}$ is induced by
   an electric field ${\vec E}$, and the magnetization ${\vec M}$ is
   induced by a magnetic field ${\vec H}$, one needs a
   set of equations to describe these processes. The equations are given,
 e.g., in \cite {Ruppin} as:
  \ber
  \ddot{\vec P} +\Gamma _e \dot {\vec P} + \omega^2_r {\vec P} =
  \epsilon_0 \omega^2_p {\vec E}
  \label {eleom}\\
  \ddot{\vec M} +\Gamma _h \dot{\vec M} + \omega^2_0 {\vec M} =
  F \omega^2_0 {\vec H}
  \label {mageom}
  \enr
 with all symbols and notation as defined in \cite {Ruppin}.
In addition, one has the Maxwell-equations given in the previous sections and
 derived from the Lagrangean of \er{staticlag}
using the contravariant four-vectors $A^\alpha= (\vec A,\Phi)$.

Turning now to dissipative cases, following \cite{Hannay}
 -\cite{Bekenstein1982} we introduce a set of
new fields, here named canonical fields, which are formally distinguished from
 the previous, physical fields by writing them in lower case  symbols.
    These fields (assumed to be real) are  made to be part of  a Hamiltonian or
  Lagrangean formulation, in contrast to the
preceding physical quantities (written in upper case symbols),
with which this cannot be done \cite{Hannay}. Explicitly, we shall work with
 the scaled polarization $p^{\alpha}$ and magnetization $ m^{\alpha}$;
 the electromagnetic fields
$e^{\alpha}$, $h^{\alpha}$, the (contravariant) vector-potentials $a^{\alpha}$,
 and (again following \cite {Hannay}) also
the fields  $r^{\alpha}, n^{\alpha}$ conjugate to $ p^{\alpha}$ and to
$ m^{\alpha}$ (as defined below). The choice
of the fields (and of the various constant, scaling factors) is guided by the
 requirement that we recapture the form of the equations of motion and the energy
  density currently widely
 employed in solid state optics in the appropriate limiting case of no dissipation.
 (This requirement is at times detrimental to the formal symmetry between
 the modes linked with the polarization and the magnetization,
 respectively.)  The physical meaning of the canonical fields will be made
 clear later by comparing
their equations of motion with those of the physical fields.
(For a similar procedure see \cite{Hannay}.) In addition, we shall introduce
 two new scalar fields: first $\Psi$, representing a
  degree of freedom associated with the dissipation of the
 polarization (analogous to the degree of freedom denoted with the same
  symbol in \cite {Bekenstein1982}) and, secondly, the "sink-field"
  $\Phi$ for the dissipation in the  magnetic mode.

  \section {The dissipative Lagrangean}

  This has the following parts:
  \beq
  {\cal L} = {\cal L}_{eb}+ {\cal L}_{pe} +  {\cal L}_{mh} +
  {\cal L}_{pr} +  {\cal L}_{mn} +
  {\cal L}_{\Psi} + {\cal L}_{\Phi} + {\cal L}_{pr-\Psi} + {\cal L}_{mn-\Phi}
    \label {Lagrangean}
  \enq
  This Lagrangean contains first the Lagrangean of \er{staticlag},
there written in terms of the electromagnetic fields, and now re-expressed
in the new lower-case variables as
  ${\cal L}_{eb}+ {\cal L}_{pe} + {\cal L}_{mh}$.
  In addition, the above expression contains the Lagrangeans in the polarization modes,
  the magnetization modes, the electromagnetic fields, the $\Psi$ and
  $\Phi$ sink fields and then the interaction-Lagrangeans between
  polarization and electric field, magnetization and magnetic
  field and, finally, the polarization and magnetization with their
  respective sink fields.
  Writing these out fully:
  \ber
  {\cal L}_{eb} &=&  \frac{1}{2}[\epsilon_0 e_1^2 - \frac{b_1^2}{\mu_0}]\\
  {\cal L}_{pe} &=&  p e_1\\\
  {\cal L}_{mh} &=& \mu_0 m h_1\\
  {\cal L}_{pr} &=&\frac{1}{2\epsilon_0\omega^2_p}(2\dot{p}r -\omega^2_r p^2
   - r^2)
    \label {lagrfirst}\\
  {\cal L}_{mn} &=&\frac{\mu_0}{2F\omega^2_0}(2\dot{m}n -\omega^2_0 m^2
   - n^2)     \\
    {\cal L}_{\Psi} &=& {1\over{2}}\kappa{\dot{\Psi}}^2 - {1\over{2}}\rho\Psi^2
   \label{LagrPsi}\\
  {\cal L}_{\Phi} &=& {1\over{2}}\lambda{\dot{\Phi}}^2 - {1\over{2}}\nu\Phi^2
\label {LagrPhi}\\
  {\cal L}_{pr-\Psi} &=& -\frac{1}{\epsilon_0 \omega^2_p}pr\dot{\Psi} \\
  {\cal L}_{mn-\Phi} &=&-\frac{\mu_0}{F \omega^2_0}mn\dot{\Phi}
  \label {Lagrlast}
  \enr
 having chosen the polarization and magnetization vectors to be along the
 $x$ (or 1) axis. Physically, the new degrees of freedom $\Psi$ and $\Phi$
  can be associated with
 some sort of relaxation mechanism for the polarization and the magnetization,
 respectively. In the above, we have chosen for  $ {\cal L}_{\Psi}$
  and ${\cal L}_{\Phi}$ what appear to us
 the simplest type of expressions that can represent unspecified (and,  so
 far,  arbitrary) degrees of freedom. The symbols $\kappa,\rho,\lambda,\nu$
 stand for constant, non-negative coefficients, whose values
 depend on the nature of the relaxation mechanisms.

 Equating to zero the variation of $\cal L$ with respect to each degree of
  freedom leads to the
  equations of motion, in accordance with the Euler-Lagrange equations.
  For the variables $r$ and $n$ the equations obtained are:
\ber
r &=& {\dot p} - p{\dot{\Psi}}
  \\
n &=& {\dot m}-m{\dot{\Phi}}
\enr
which can be inserted into the equations for $p$ and $m$. These take now the
following form:
  \ber
    \ddot{p}-( \ddot{\Psi}+{\dot{\Psi}}^2 -\omega^2_r)p & = &\epsilon_0 \omega^2_p
    e_1
    \label{peom}\\
    \ddot{m}-( \ddot{\Phi}+{\dot{\Phi}}^2 -\omega^2_0)p & = &F \omega^2_0
    h_1
    \label{meom}
    \enr
The electromagnetic equations of motion are identical to Maxwell's equations
in the new (small letter) variables. Finally, for the sink coordinates one
 has
\ber
\ddot {\Psi}+\frac{\rho}{\kappa}\Psi =
{1\over{\kappa\epsilon_0\omega^2_p}}\frac{\partial(pr)}{\partial t}
\label {Psieom}\\
\ddot {\Phi}+\frac{\nu}{\lambda}\Phi =
\frac{\mu_0}{\lambda\omega^2_0}\frac{\partial(mn)}{\partial t}
\label {Phieom}
\enr

  From \er{Tkj} we obtain the expression for the energy density $T^0_0$,
 namely,
\beq
 T^0_0 =  \sum_{k}  \frac{\partial {\cal L}}{\partial \dot{u_{k}}}\dot{u_{k}}-
  {\cal L}
  \label{edk}
  \enq
where the sum is  over all degrees of freedom. We separate  the
electromagnetic, the polarization and the magnetization  parts
\beq
T^0_0=(T^0_0)_{e,h}+(T^0_0)_{p,\Psi} +(T^0_0)_{m,\Phi}
\label {separate}
\enq
where the first term has the well known form of the electromagnetic energy density
given in \er{edstatic} and repeated here for completeness:
\beq
(T^0_0)_{e,h}= \frac{\epsilon_0}{2} e^2+\frac{\mu_0}{2} h^2 + \frac{\mu_0}{2} m^2
\label{emden}
\enq
For the others we obtain
\ber
(T^0_0)_{p,\Psi} & = & {1\over{2\epsilon_0 \omega^2_p}}\big[({\dot {p}-
p{\dot{\Psi}}})^2
+\omega^2_r p^2\big]  +    \frac{\kappa}{2}{\dot{\Psi}}^2+
    \frac{\rho}{2}\Psi^2\\
    & = &{1\over{2\epsilon_0 \omega^2_p}}({\dot {p}}^2
+\omega^2_r p^2)  +    \frac{\kappa}{2}{\dot{\Psi}}^2+
    \frac{\rho}{2}\Psi^2 + {1\over{2\epsilon_0
    \omega^2_p}}p^2{\dot{\Psi}}^2\
    \nonumber\\
    & - & {1\over{2\epsilon_0
    \omega^2_p}}\frac{\partial {p^2}}{\partial t}{\dot{\Psi}}
     \label {edp2}
 \enr
 and
\ber
(T^0_0)_{m,\Phi} & = & \frac{\mu_0}{2F\omega^2_0}\big[(\dot
{m}-m{\dot{\Phi}})^2+{\omega^2_0}m^2\big]
        + \frac{\lambda}{2}{\dot{\Phi}}^2 + \frac{\nu}{2}{\Phi}^2\\
& = &\frac{\mu_0}{2F \omega^2_0}({\dot {m}}^2
+\omega^2_0 m^2)  +    \frac{\lambda}{2}{\dot{\Phi}}^2+
    \frac{\nu}{2}\Phi^2 + \frac{\mu_0}{2F\omega^2_0}m^2{\dot{\Phi}}^2\
    \nonumber\\
    & - & \frac{\mu_0}{2F\omega^2_0}\frac{\partial {m^2}}
    {\partial t}{\dot{\Phi}}
\label{edm2}
\enr
The first-written form (=sum of squares with non-negative coefficients) of
either quantity guarantees that
 each part of the energy density is positive (non-negative). The usual
  expressions for the energy density, e.g. in \cite{Jackson} or
\cite {Ruppin}, differ from the above by the presence of the terms in
$\Phi$ and $\Psi$ and their time derivatives. In the following section it
 is our purpose to
eliminate these variables by making use of the equations of motion,
\er{Psieom} and \er{Phieom}.

\section{A solvable case}

We take a  simplified
case when the coefficients $\rho$ and $\nu$ are both zero. Then the
equation of motion for the polarization sink-mode
\er{Psieom} can be integrated. (A similar procedure applies to the
magnetization sink-variable.) We assume the following initial
 conditions for $\Psi(t)$
 \beq
 \Psi(0)=0, ~~~~\dot{\Psi}(0)=\Gamma_e /2
 \label{ic}
 \enq
 The reason for these choices is that with them for short times,
 $t<<\frac{2}{\Gamma_e}$ the differential  \er{eleom} is regained. (This will be
 shown presently. Cf. also \cite {Hannay}.) Then from \er {Psieom}
 \beq
 \ddot{\Psi}={1\over{2\kappa\epsilon_0 \omega^2_p}}\frac{{\partial}^2}{ \partial
 t^2}{p^2(t)}-{1\over{\kappa\epsilon_0 \omega^2_p}}\frac{\partial ({p^2(t)}\dot{\Psi})}{\partial t}
\label{Psiddot}
\enq
Integrating once
\beq
\dot{\Psi}={1\over{2\kappa\epsilon_0 \omega^2_p}}\frac{\partial{p^2(t)}}{ \partial
 t}- \frac{{p^2(t)}\dot{\Psi}}{\kappa\epsilon_0 \omega^2_p}+C
 \label{Psidot}
 \enq
 leading to
 \beq
 (1+\frac{{p^2}(t)}{\kappa\epsilon_0 \omega^2_p}){\dot{\Psi}}={1\over{2\kappa
\epsilon_0 \omega^2_p}}\frac{\partial {p^2(t)}}{ \partial t} + C
 \label{Psidot2}
 \enq
with the (first) integration constant given by
\beq
C=\frac{\Gamma_e}{2} (1+\frac{{p^2(0)}}{\kappa\epsilon_0 \omega^2_p})-
 \frac{p(0){\dot{p}(0)}}{\kappa \epsilon_0\omega^2_p}
\label{C}
\enq
Integrating once more and arranging for satisfaction of the first
initial condition in \er{ic}, we finally obtain:
\ber
\Psi(t)& = & {1\over{2}}\ln\big[ \frac{ (1+\frac{{p^2}(t)}{\kappa\epsilon_0
\omega^2_p})}{ (1+\frac{{p^2}(0)}{\kappa \epsilon_0\omega^2_p})}\big] \nonumber
\\ &  & + \big[\frac{\Gamma_e}{2} (1+\frac{{p^2}(0)}{\kappa \epsilon_0\omega^2_p})-
\frac{p(0){\dot{p}(0)}}{\kappa \epsilon_0 \omega^2_p}\big] \int_0^t\frac{dt'}
{ 1+\frac{{p^2}(t')}{\kappa\epsilon_0\omega^2_p}}
\label{Psif}
\enr
When this expression and \er{Psidot} are substituted into \er{edp2}, one
obtains after  considerable simplification the following expression for
the energy density arising from the time varying "canonical"
polarization $p(t)$:
\beq
(T^0_0)_{p,\Psi} =  \frac{1}{2\epsilon_0 \omega_p^2}\Bigl(\frac{{\dot
p}^2(t)+ \kappa \epsilon_0 \omega^2_p C^2}{1+\frac{p^2(t)}{\kappa\epsilon_0\omega^2_p}} +\omega_r^2 p^2(t)\Bigr)
\label{ped3}
\enq
When we assume that $\Gamma_e$ is small (precisely, $\Gamma_e \sqrt \kappa <<1$),
and so is $\dot p(0)$ (this will be confirmed in the next section),
 then  $C^2$ will be
a second order correction which may be neglected:
\beq
(T^0_0)_{p,\Psi} =  \frac{1}{2\epsilon_0 \omega_p^2}\Bigl(\frac{{\dot
p}^2(t)}{1+\frac{p^2(t)}{\kappa\epsilon_0\omega^2_p}} +\omega_r^2 p^2(t)\Bigr)
\label{ped3b}
\enq
A similar expression is obtained for the part of the energy density
involving the "canonical" magnetization $m(t)$, the variables $n(t)$
and $\Phi(t)$ having been eliminated through their equations of motion,
\beq
(T^0_0)_{m,\Phi} =  \frac{\mu_0}{2F\omega_0^2}\Bigl(\frac{{\dot
m}^2(t)}{1+\frac{\mu_0 m^2(t)}{\lambda F \omega^2_0}} +\omega_0^2 m^2(t)\Bigr)
\label{med3}
\enq
The above expressions, \er {ped3b} and \er {med3}, are quite similar to those in equation (11) of
 Ruppin \cite{Ruppin},
 except that they are written in the canonical (small letter) variables, rather than
 in the physical variables
(for  relations between these, see immediately below), and that they contain
 time dependent denominators.

The main results of this work, \er {ped3} and its analogue for the magnetization
energy density,  are exact and contain nonperturbative corrections
to the energy density, due to the presence of the sink degrees of freedom.
While exact, they are model dependent in the sense that sinks represented by
different Lagrangeans would lead to different energy  densities. This is
 clear, due to
the presence in the energy densities of the parameters $\kappa$ and $\lambda$
 that were introduced
 in the Lagrangean in \er{LagrPsi} and \er{LagrPhi}.
It is of interest to note that the non-dissipative limit is not regained
when $\Gamma_e,\Gamma_h \to 0 $, but only when {\it  also} $\kappa, \lambda
\to \infty$.

\section{ The physical fields}

To recapture the basic equations of motion \er{eleom} and \er{mageom} for
 the physical polarization variable, we proceed as follows: We postulate
\ber
p(t) & = & e^{\Psi(t)} P(t)\\
m(t) & = & e^{\Phi(t)} M(t)
\label{pPmM}
\enr
Then, from \er{ic}, for short times $0<t<< \frac{2}{\Gamma_e}$,
\beq
\Psi \approx \Psi(0) + \dot \Psi(0) t =\Gamma_e t/2
\label{initPsi}
\enq
This turns (the vector form of) \er{peom} into the following:
\beq
 \ddot{\vec P} +\Gamma _e \dot {\vec P} + \omega^2_r {\vec P} =
  \epsilon \omega^2_p e^{-\frac{\Gamma_e t}{2}}{\vec e}
  \label {neweleom}
  \enq
and likewise for the magnetization variables. Recalling \er{eleom} and
 \er {mageom},  we can thus extrapolate to later times so as to identify
\ber
{\vec e(t)} & = & e^{\Psi(t)} {\vec E(t)}\\
{\vec m(t)} & = & e^{\Phi(t)} {\vec H(t)}
\label{eEmM}
\enr
Thus, {\it all} the "canonical" variables are the fields in which the decay
 of the physical field variables has been reinstated. On the other hand, the
  decay is itself dependent
 on the fields. (Cf. \cite {Neufeld69}.) Furthermore, Maxwell's equations for the physical fields
 are also modified, just as in  {\cite {Bekenstein2002} and
  \cite {Bekenstein1982} (eq. (14) and  eq. (12), respectively).
 \section{Conclusion}
 Using standard Lagrangean formalism, we have obtained a
 unique energy density for a dissipative medium
 (capable also  of sustaining an electric polarization and/or a magnetic one).
 It has been found necessary to introduce two additional degrees of freedom
 ("sink-modes"), associated with  decay mechanisms in the electrical and magnetic
 modes. The derived contributions to the energy density, shown in \er{ped3} - \er{med3},
 are all positive and are  similar in their form to the corresponding results
 in \cite {Ruppin}. They contain particular physical parameters, as anticipated in
\cite {Loudon}. The energy density is part of a conservation
 equation involving also an energy-current (momentum) density (the Poynting
 vector)  formally given by the tensor-components $T^a_0 (a=1,2,3)$
 shown in \er {Tkj}. These properties appear to be true in general,
 namely, for a large variety of systems that are dissipative.

  A solvable model has been worked out in this paper. Both in this case
 and in a general one, the energy density has corrections not
 appearing in other approaches not based on a Lagrangean. On the other
 hand, in a spatially homogeneous infinite medium (such as treated in
  this paper) the momentum density is still the Poynting vector.
   This can be seen from \er{Tkj}, since the second term in that expression is
  absent (for $\alpha$ differing from $\beta$) and there are no space derivatives
  in the Lagrangean, other than in the electromagnetic part. (As a
   consequence, the Lagrangean is not fully Lorentz-invariant.)

    For a medium possessing space varying properties (including, under certain
   circumstances, a finiteness of size) the procedure outlined in this paper
  can be extended to space-coordinates. This means the introduction
  of new, space-dependent "sink" degrees of motion (e.g., at the
  boundaries). As a consequence, additional terms would appear also in
  the momentum density. Apart from an interesting extension of the Poynting
  vector concept, the suggested approach can also have the practical
  use of solving electromagnetic problems involving dissipation, such
  as laser gain and loss calculations, microwave losses in wave guides
 and  cavities and transformer performance. The approach can be further
  extended to other fields of macroscopic physics such as viscous fluid
 dynamics. Work in these directions is under progress.

\end{document}